\documentclass[twocolumn]{aastex7}
\usepackage{color}
\usepackage[titletoc]{appendix}
\usepackage{amsmath}
\usepackage{amssymb}
\usepackage{mathtools}
\usepackage{upgreek}
\usepackage{comment}
\usepackage{enumitem}
\usepackage{gensymb}
\usepackage{natbib}
\usepackage{graphicx}
\usepackage{bm}
\usepackage{totcount}
\usepackage{multirow}
\usepackage{hyperref}
\usepackage{cleveref}
\usepackage{tabularx}
\usepackage[T1]{fontenc}

\newcommand{\tc}{\tilde{c}}

\newtotcounter{citnum} 
\def\oldbibitem{} \let\oldbibitem=\bibitem
\def\bibitem{\stepcounter{citnum}\oldbibitem}

\crefname{subsection}{subsection}{subsections}

\shortauthors{}
\shorttitle{}

\begin{document} 

\title{Case Study of a Super-eccentric Warm Jupiter Migrating via Equilibrium and Dynamical Tides}

\author[0009-0003-3584-6698]{Donald Liveoak}
\affil{Department of Physics, University of Michigan, Ann Arbor, MI 48109}
\affiliation{Department of Physics, Massachusetts Institute of Technology, Cambridge, MA 02139, USA}
\affiliation{MIT Kavli Institute for Astrophysics and Space Research, Massachusetts Institute of Technology, Cambridge, MA 02139, USA}
\email{dliveoak@umich.edu}

\author[0000-0003-3130-2282]{Sarah C. Millholland}
\affiliation{Department of Physics, Massachusetts Institute of Technology, Cambridge, MA 02139, USA}
\affiliation{MIT Kavli Institute for Astrophysics and Space Research, Massachusetts Institute of Technology, Cambridge, MA 02139, USA}
\email{sarah.millholland@mit.edu}

\author[0000-0002-3752-3038]{Michelle Vick}
\affiliation{Department of Physics, Brown University, Providence, RI 02912, USA}
\email{michelle_vick@brown.edu}

\begin{abstract}
A leading theory for hot Jupiter formation is high-eccentricity migration, in which planets are born at large separations and excited to near-unity eccentricities, creating extreme tidal dissipation that shrinks and circularizes their orbits. The most direct evidence for this scenario is the detection of highly eccentric planets caught in the act of migrating. Although such planets are rare, the recent discovery of TIC~241249530~b -- a super-eccentric ($e = 0.94$) and retrograde hot Jupiter progenitor -- presents an ideal case study to explore high-eccentricity migration and the accompanying tidal physics. In this paper, we examine the migration history of TIC~241249530~b and test two different models of tidal dissipation: equilibrium tides and chaotic dynamical tides. We show that TIC~241249530~b's properties can be explained by high-eccentricity migration triggered by von Zeipel-Lidov-Kozai oscillations induced by the observed distant binary star in the system, but only if the dominant tidal dissipation takes the form of equilibrium tides. Chaotic dynamical tides fail to explain the system because they require the planet to have migrated substantially closer to its star.

\end{abstract}

\section{Introduction}
\label{sec: Introduction}

Hot Jupiters (HJs) are broadly defined as planets with mass $\gtrsim 0.5 M_J$ and orbital period $\lesssim 10$ days. Despite the abundance of HJs in the observational data \citep[e.g.][]{2022AJ....164...70Y} and their rich history in the early era of exoplanet science \citep{mayor1995jupiter}, their dynamical histories remain poorly understood \citep{dawson2018origins}. It is widely believed that these planets cannot form near their current orbits due to a deficiency of solid mass needed to form a planetary core \citep{rafikov2005can, rafikov2006atmospheres}.

A commonly proposed mechanism for HJ formation is high-eccentricity migration \citep[HEM;][]{2003ApJ...589..605W,
2007ApJ...669.1298F, 2011Natur.473..187N, dawson2018origins}. In this scenario, HJs form far ($\gtrsim$ a few AU) from their host star and are perturbed onto highly eccentric orbits via interactions with distant bodies. These highly eccentric warm Jupiters (WJs) have short periapse distances that lead to strong tidal dissipation which gradually shrinks and circularizes the orbit until the planet ends up as a hot Jupiter. While a plethora of pathways may exist for generating the required eccentricity excitation \citep[e.g.][]{1996Sci...274..954R, 2008ApJ...686..580C, 2011ApJ...735..109W, 2015ApJ...805...75P}, one of the chief theories is the von Zeipel–Lidov–Kozai (vZLK) mechanism \citep{1910AN....183..345V, 1962AJ.....67..591K, 1962P&SS....9..719L}.

The vZLK effect occurs when a planet's orbit is perturbed by a stellar or planetary companion with an initially large mutual inclination \citep{2013MNRAS.431.2155N, 2016ARA&A..54..441N, 2016ApJ...829..132P}. This perturbation results in quasi-periodic exchanges between the planet's eccentricity and inclination, known as vZLK cycles or vZLK oscillations. During the high-eccentricity phases of the vZLK oscillations, the planet's pericenter distance can shrink by several orders of magnitude to a regime where short-range forces (SRFs), including tidal forces and general relativity become relevant and eventually decouple the orbit of the planet from the companion \citep{2015MNRAS.447..747L}. Inward migration is driven by the dissipation of orbital energy due to tides in the high-eccentricity phases of the vZLK oscillations \citep{munoz2016formation}. HEM via vZLK oscillations is successful in describing various aspects of the hot Jupiter population, including the orbital period distribution, stellar spin-orbit misalignments, and presence of long period perturbers \citep[e.g.][]{2016MNRAS.456.3671A, hamers2017formation, 2018AJ....156..128S, vick2023high, 2025ApJ...980L..31W}.  

Despite its successful demonstration in simulations, the relative paucity of highly eccentric gas giants in the observational data presents a significant challenge for this formation channel \citep{2012ApJ...750..106S, 2015ApJ...798...66D, 2023AJ....165...82J, gupta2024hot}. One possible resolution to this issue is that the planets move quickly through their most eccentric phase due to exceptionally efficient dissipation \citep{vick2019chaotic}. Investigating this possibility requires understanding the precise physics of tidal dissipation within migrating giant planets, which is still uncertain. 

Most studies of HEM have considered the equilibrium tide (ET) model in the weak friction approximation, in which the planet develops a hydrostatic distortion with a time lag \citep{1998ApJ...499..853E, 2001ApJ...562.1012E}. The simple analytics of this theory of tides have formed the basis of models of HEM \citep[e.g.][]{fabrycky2007shrinking, 2011ApJ...735..109W}. An alternative mechanism that is likely especially relevant at high eccentricities is the dynamical tide (DT) model, which considers dissipation through vibrational modes within the interiors of giant planets \citep{2018AJ....155..118W, 2018MNRAS.476..482V}. At extreme eccentricities, DTs excited within WJs are thought to behave chaotically and rapidly dissipate energy, thus leading to fast migration. This process is known as chaotic dynamical tides (CDTs), and has been explored in several high-eccentricity migrations scenarios \citep{vick2019chaotic, teyssandier2019formation}. In particular, dynamical tides have gained recent traction for their potential role in the migration and mass loss of hot Jupiters \citep[e.g.][]{weldon2026saving, 2026arXiv260620789Z}. We stress that even if CDTs operate in the high-eccentricity regime, ETs are necessary in order for the WJ to complete migration to become a HJ.

Testing these two models of tides -- ET vs. CDT -- is ideally suited to the study of individual systems currently going through the highly eccentric phase. The recent discovery of TIC~241249530~b, a super-eccentric ($e\approx 0.94$) WJ with an observed stellar binary companion, provides an excellent case study opportunity \citep{gupta2024hot}. In this paper, we analyze both tidal migration scenarios with the goal of assessing which is more likely responsible for TIC~241249530~b's observed orbit, and we constrain TIC~241249530~b's dynamical history. Here we do not attempt to couple the orbital migration with tidally-induced interior structure changes \citep{2026ApJ...997..138H,
2026ApJ...997..139H, 2026arXiv260620789Z}, but later in the paper we will show that this approach is warranted for this particular planet. We first give an overview of the system (Section \ref{sec:TIC}). We then study its formation from the framework of ETs (Section \ref{sec:ET}) and CDTs (Section \ref{sec: DT}). In Section \ref{sec: discussion}, we discuss planetary structure evolution relevant to our calculations and attempt to draw conclusions for the broader population of warm Jupiters and hot Jupiters. 

\section{Overview of TIC 241249530} \label{sec:TIC}

\begin{table}[ht]
\centering
\caption{System parameters of TIC 241249530.}
\label{tab:TIC parameters}
\begin{tabular}{lc}
\hline
\hline
Parameter & Value \\
\hline
\multicolumn{2}{c}{\textit{Host Star Parameters}} \\
\hline
Stellar mass, $M_\star$ & $1.271^{+0.061}_{-0.068}\,M_\odot$ \\
Stellar radius, $R_\star$ & $1.397^{+0.025}_{-0.028}\,R_\odot$ \\
\hline
\multicolumn{2}{c}{\textit{Planet Parameters}} \\
\hline
Planet mass, $M_p$ & $4.98^{+0.16}_{-0.18}\,M_{\rm J}$ \\
Planet radius, $R_p$ & $1.186^{+0.037}_{-0.040}\,R_{\rm J}$ \\
Orbital period, $P$ & $165.77190^{+0.00027}_{-0.00028}\,\mathrm{days}$ \\
Semi-major axis, $a$ & $0.641^{+0.010}_{-0.012}\,\mathrm{AU}$ \\
Eccentricity, $e$ & $0.9412^{+0.0009}_{-0.0009}$ \\
Argument of periastron, $\omega$ & ${42.32^\circ}^{+0.40^\circ}_{-0.36^\circ}$ \\
Projected spin-orbit obliquity, $\lambda$ & ${163.5^\circ}^{+9.4^\circ}_{-7.7^\circ}$ \\
\hline
\multicolumn{2}{c}{\textit{Binary Parameters}} \\
\hline
Stellar mass, $M_{\star}$ & $0.453^{+0.012}_{-0.012}\,M_{\odot}$\\
Projected separation & $ 1664\pm11$ AU \\
\hline
\hline
\end{tabular}
\end{table}

TIC 241249530 is an F-type star around which a single transit was first detected with the Transiting Exoplanet Survey Satellite (TESS) \citep{gupta2024hot}. Radial velocity follow-up confirmed the planet, TIC~241249530~b, to be a massive WJ ($\sim5 \ M_{\mathrm{Jup}}$ on a $\sim165$ d orbit) with a record setting eccentricity of $0.94$. The system also hosts a distant binary star with an uncertain orbit but a projected separation of $1664 \pm 11$ AU. The basic parameters of the system are listed in Table \ref{tab:TIC parameters}. The system age is estimated to be about $3.2\pm0.5$ Gyr. 

\cite{gupta2024hot} showed that the parameters of the system are consistent with the planet having undergone vZLK oscillations induced by the perturbations from the binary star, which then triggered high eccentricity migration. They estimated that the planet exited from its vZLK oscillations $\sim200$ Myr ago and is undergoing isolated tidal evolution towards becoming a hot Jupiter. However, their calculations only considered ET and presented a single integration with a proof of concept of the possible history. In the next sections, we aim to build upon their work by considering both ET and DT and exploring a range of possible initial conditions. We will examine whether the data favors ET or DT for the planet's migration history.

\section{HEM with Equilibrium Tides} \label{sec:ET}

\subsection{Octupole vZLK oscillations with short-range forces}

To analyze the dynamical history of TIC~241249530~b in the context of vZLK migration, we work under the secular and hierarchical approximations, assuming that the perturbing body is much more massive and on a substantially wider orbit than the planet. Given these assumptions, one can expand to the octupole order in $a/a_b$, where $a$ and $a_b$ are the semi-major axes of the planet and perturbing companion, to find approximate equations of motion for the planet's orbital elements \citep{lithwick2011eccentric, katz2011long, 2013MNRAS.431.2155N}. \cite{liu2015suppression} augments these equations of motion to include the effects of ETs and general relativistic precession, which have the effect of limiting the maximum eccentricity $e_{\text{max}}$ of the planet during vZLK oscillations. In this section, we neglect DTs and isolate the effect of ETs.

We analyze the dynamical history of TIC~241249530~b using the characteristic timescale for vZLK migration driven by ETs \citep{2016MNRAS.456.3671A, vick2019chaotic}:
\begin{align*}
    t_{\text{mig}}^{-1} &= \frac{1.27}{\text{Gyr}} \left(\frac{k_{2}}{0.37}\right)\left(\frac{\Delta t}{1 \,\text{s}}\right)\left(\frac{M_\star}{M_\odot}\right)^2\left(\frac{M_p}{M_J}\right)^{-1}\\
    &\times \left(\frac{R_p}{R_J}\right)^5\left(\frac{a_0}{1\,\text{au}}\right)^{-1}\left(\frac{a_0(1-e_\text{max,0})}{0.025\, \text{au}}\right)^{-7},
\end{align*}
where $k_2$ is the Love number, $\Delta t$ is the tidal time lag, $a_0$ is the initial semi-major axis of the planet, and $e_{\max, 0}$ is the maximum of the eccentricity during the first vZLK oscillation. 
\begin{figure}
    \centering
    \includegraphics[width=\linewidth]{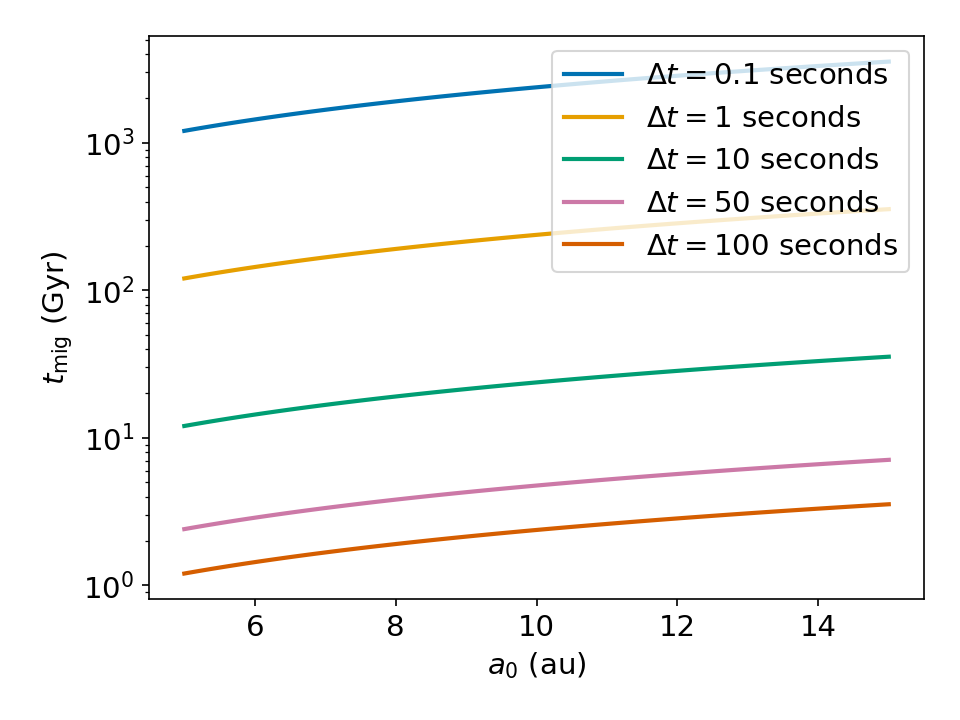}
    \caption{Migration timescale $t_{\text{mig}}$ vs. initial semi-major axis $a_0$ for TIC~241249530~b considering multiple possible tidal time lags.}
    \label{fig:migration-timescales}
\end{figure}

\Cref{fig:migration-timescales} depicts the migration timescales for TIC~241249530~b, assuming $k_2=0.25$ and angular momentum is conserved in the post-vZLK dynamics so that $e_{\text{max,0}}$ can be calculated via the reduced angular momentum
\begin{equation*}
    j^2 \equiv a(1-e^2) = a_0(1-e_{\text{max,0}}^2).
\end{equation*}
We find that in order for TIC~241249530~b to have migrated to its current orbit within the inferred age of the system ($\sim$ a few Gyr), we must have ${\Delta t \gtrsim 10 \, \text{seconds}}$, which is substantially more dissipative than the ${\Delta t_J \approx0.1}$ seconds of Jupiter \citep{leconte2010tidal, 2010ApJ...723..285H, 2012arXiv1209.5724S}. We note that this value is consistent with previous studies on HJ formation via HEM, which generally require $\Delta t \sim 1-10$ seconds \citep{2012arXiv1209.5724S, 2015ApJ...799...27P, 2015ApJ...805...75P}.

\subsection{Application to TIC~241249530~b's formation}

\begin{figure}
    \centering
    \includegraphics[width=\linewidth]{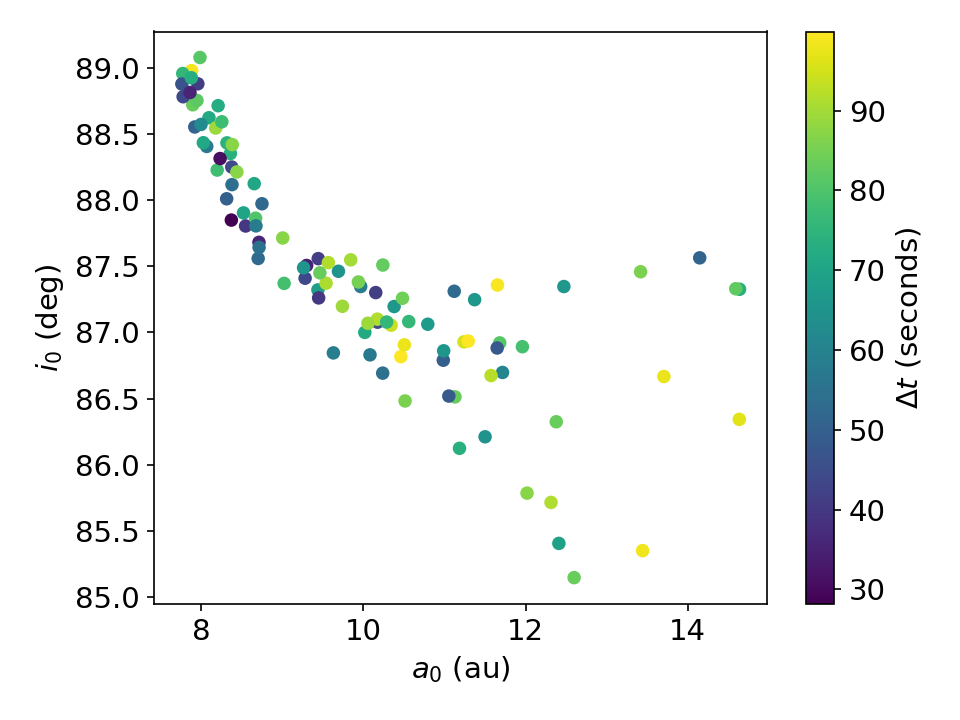}
    \caption{Initial mutual inclination vs. initial proto-WJ semi-major axis for which $j^2$ is within three standard deviations of the observed value for systems in \texttt{suite A} (in which $a_b$ and $e_b$ are fixed). The color of each point is determined by the proto-WJ's time lag $\Delta t$.}
    \label{fig:parameter-space}
\end{figure}

\begin{figure}
    \centering
    \includegraphics[width=\linewidth]{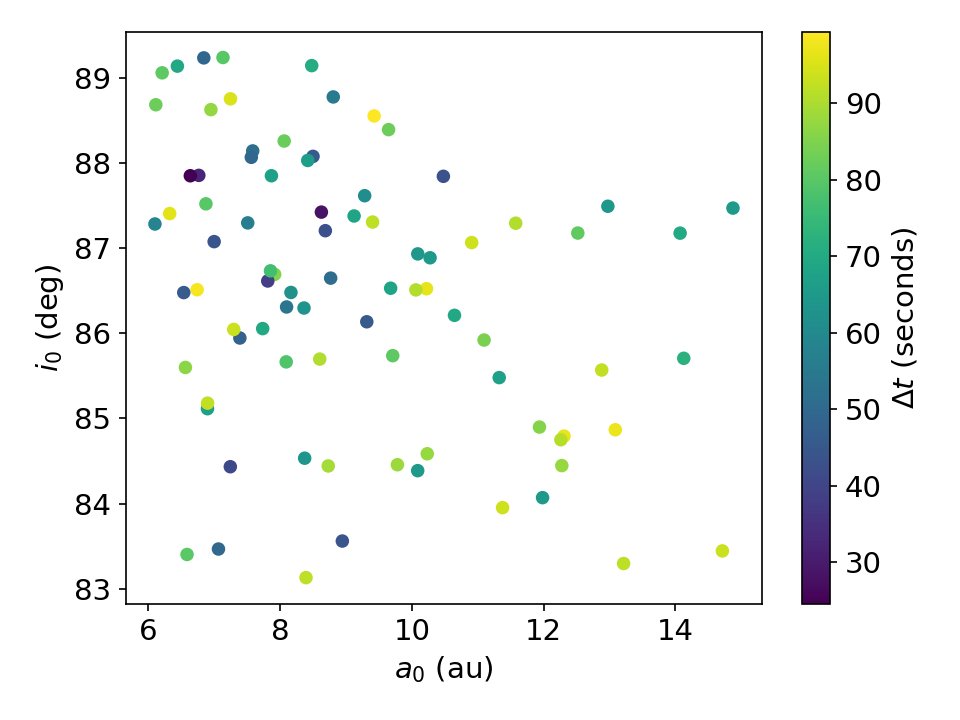}
    \caption{Same as \Cref{fig:parameter-space}, except for \texttt{suite B} (in which $a_b$ and $e_b$ are varied). Compared to the case where the binary parameters are fixed, a larger region of $(a_0,i_0)$ for the proto-WJ becomes consistent with the observed parameters of TIC~241249530~b.}
    \label{fig:parameter-space-binary-params}
\end{figure}

\begin{figure*}
    \centering
    \includegraphics[width=\linewidth]{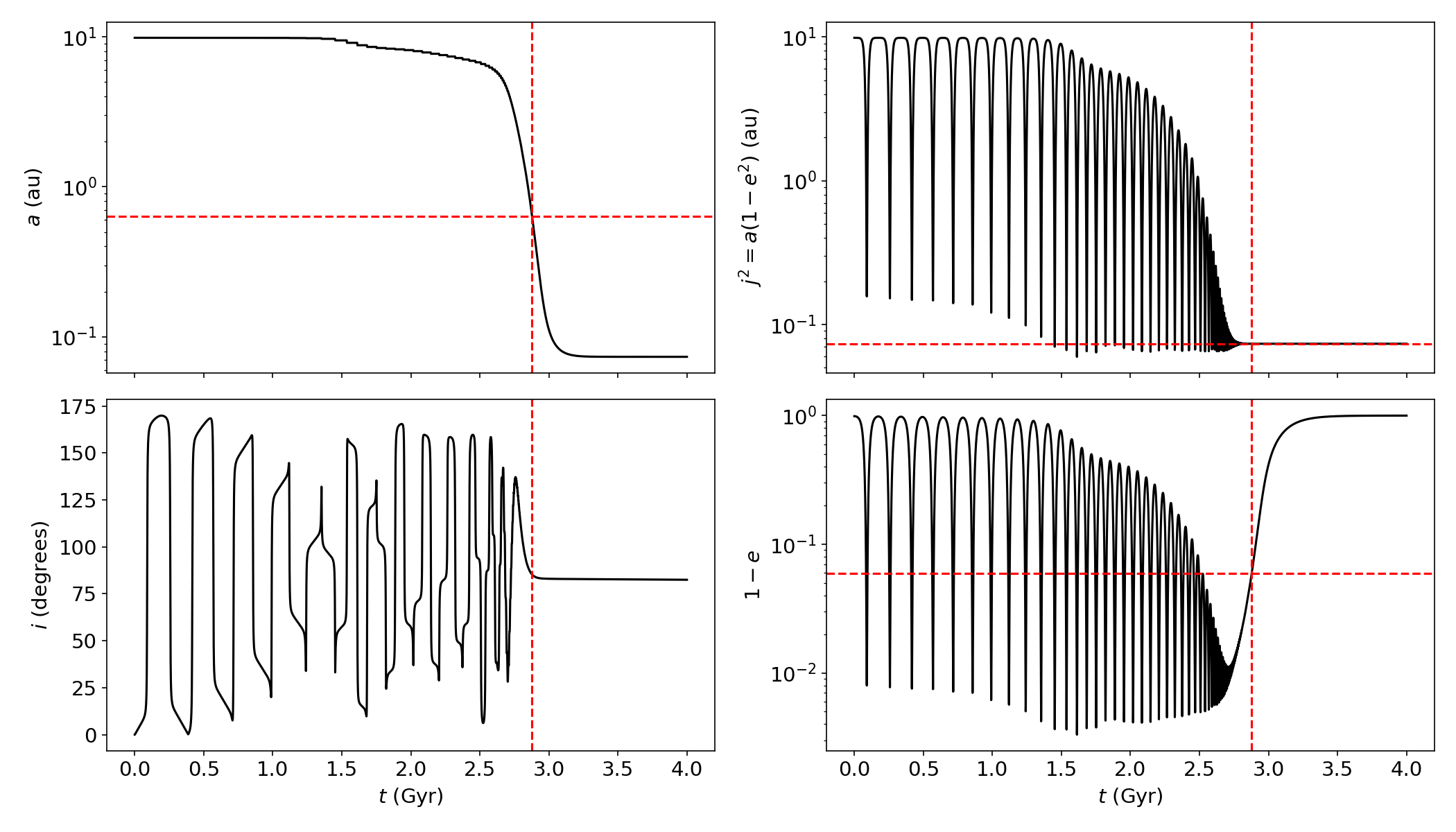}
    \caption{Example of the secular evolution of a system from \texttt{suite A} that matches the observed orbital parameters of TIC~241249530~b. In this case, $a_0 = 9.86\, \text{au}$ and $i_0 = 85.03$ degrees. The dashed red lines correspond to $t=2.88$ Gyr, where $a=0.638$ and $e=0.941$. For this system, $\Delta t=45.1$ seconds.}
    \label{fig:secular-sim-tic}
\end{figure*}

We perform orbital integrations in the secular and hierarchical approximations by utilizing the publicly available \texttt{SecularMultiple} code developed by \cite{2016MNRAS.459.2827H} and extended in \cite{2018MNRAS.476.4139H, 2020MNRAS.494.5492H}. This code integrates the long-term dynamical evolution of systems with any number of bodies in a hierarchical structure of nested binaries. The Hamiltonian is expanded in terms of ratios of the separations of all binaries in the
system and then averaged over all orbits
in the system (although \citealt{2020MNRAS.494.5492H} extended the code to allow some orbits to be integrated directly). The code also accounts for nodal precession as a result of equilibrium tides and 1PN general relativity, as well as tidal friction using the constant time lag approximation in the framework of \cite{1998ApJ...499..853E}. 

We carry out two integration suites, which we label \texttt{suite A} and \texttt{suite B}. In \texttt{suite A}, we vary the initial orbital and tidal parameters of the proto-WJ but leave the orbital parameters of the binary star fixed; in \texttt{suite B}, we also allow the orbital parameters of the binary star to vary.

For both integration suites, we select the mass and radius of the host star and proto-WJ consistent with their observed values (\Cref{tab:TIC parameters}). We initialize the proto-WJ on a nearly-circular orbit ($e_0=0.1$) and uniformly sample the semi-major axis $a_0$ from $6$ to $15$ au, consistent with the calculations of \cite{gupta2024hot}. We choose the tidal Love number of the proto-WJ to be $k_2=0.25$ and the tidal gyration radius to be $0.25$. We uniformly sample the tidal time lag $\Delta t $ between $10$ and $100$ seconds, which is $\sim10^2 -10^3$ times more dissipative than Jupiter \citep{leconte2010tidal}, since we find that efficient migration of the proto-WJ to the current orbit of TIC~241249530~b within $\sim$ a few Gyr requires $\Delta t \gtrsim 10$ seconds. 

As for the binary parameters, the mass of the binary is chosen according to \Cref{tab:TIC parameters}. We uniformly sample the initial mutual inclination between the proto-WJ and binary companion orbits, $i_0$, from $83^{\circ}$ to $89.5^{\circ}$; we note that the maximum eccentricity of the proto-WJ during vZLK cycles strongly depends on $i_0$ \citep{liu2015suppression}; for $i_0 \lesssim 83^{\circ}$, the maximum eccentricity of the proto-WJ is not high enough to efficiently dissipate orbital energy and quench the vZLK cycles within $\sim 10$ Gyr. We sample the argument of periapsis $\omega$ and the longitude of ascending node $\Omega$ uniformly for both the proto-WJ and the binary companion. In \texttt{suite A}, we fix the semi-major axis of the binary companion's orbit to be $a_b =1664$ au and $e_b=0.4$. In \texttt{suite B}, we uniformly sample $a_b \in [800\,\text{au}, 2000\,\text{au}]$ and $e_b \in [0, 0.8]$. Each suite consists of 4000 integrations of $4$ Gyr with a timestep of $800$ kyr.

For each simulated system in which the vZLK cycles are quenched, we compute the post-vZLK reduced angular momentum of the proto-WJ, $j^2 = a(1-e^2)$. After the planet is decoupled from the binary due to the SRFs, its angular momentum is conserved, so we expect $j^2$ to be constant after the final high-eccentricity vZLK epoch until eventually $a_{\mathrm{final}}=j^2$ on the planet's final (circular) orbit. Thus, systems whose reduced angular momentum is $j_{\text{obs}}^2 \approx a_\text{obs}(1-e_\text{obs}^2) = 0.0745\pm 0.0017\, \text{au} $, with $a_\text{obs}$ and $e_\text{obs}$ the observed semi-major axis and eccentricity, will have orbits consistent with TIC~241249530~b at some time in their evolution after vZLK is quenched.

Initial system configurations that yield WJs on final orbits with $j^2$ within three standard deviations of the observed value are shown for \texttt{suite A} in \Cref{fig:parameter-space} and \texttt{suite B} in \Cref{fig:parameter-space-binary-params}. In both cases, only systems with $\Delta t \gtrsim 30$ seconds migrate to an orbit consistent with TIC~241249530~b. This value is several orders of magnitude larger than the inferred time lag for Jupiter of $\Delta t_J\approx 0.1-1$ seconds \citep{leconte2010tidal}. For systems with $j^2 \approx 0.0745 \, \text{au}$, the WJ generally reaches $a\approx 0.64 \, \text{au}$ and $e \approx 0.94$ around $1$ to $4$ Gyr of evolution, consistent with the inferred age of TIC 241249530 from \cite{gupta2024hot}, $3.2\pm0.5$ Gyr. One example of such a system is depicted in \Cref{fig:secular-sim-tic}.

In \Cref{fig:parameter-space}, we note that a particular curve in $(a_0, i_0)$ corresponds to $j^2 \approx 0.0745 \, \text{au}$ for $a_0 \lesssim 10 \, \text{au}$; we hypothesize that this curve may be computable analytically from previous studies on vZLK migration with SRFs \citep[e.g.][]{liu2015suppression}. In \Cref{fig:parameter-space-binary-params}, we see that a wide range of initial semi-major axes and mutual inclinations may reproduce the orbit of TIC~241249530~b if the binary orbital parameters are allowed to vary.

As a result of these simulations, we conclude that vZLK oscillations combined with equilibrium tides can indeed reproduce the configuration of TIC~241249530~b, provided that the binary orbit starts nearly perpendicular with the planet's orbit and that the planet is $\sim10^2-10^3$ times more dissipative than Jupiter. This is even higher than the inferred dissipation bound, $\Delta t \sim 1-10$ seconds, found in previous HJ formation studies \citep{2012arXiv1209.5724S, 2015ApJ...799...27P}. 

We briefly comment on the feasibility of this inferred dissipation strength from theoretical grounds. \cite{2024MNRAS.527.8245L} investigated tidal dissipation mechanisms in rotating giant planets and found that inertial waves (IWs), which propagate in convective zones and are restored by the Coriolis force \citep[e.g.][]{2009MNRAS.396..794O}, are the dominant tidal dissipation mechanism whenever they are excited. IWs require the tidal period to be greater than half the planet's rotation period and yield $Q'\approx10^3(P_{\mathrm{rot}}/1\mathrm{d})^2$. \cite{2024MNRAS.527.8245L} also find that the convective damping of equilibrium tides is
much weaker than wavelike tides because, in giant planets, the convection is slow relative to the tide, which leads to a strong frequency reduction of the effective turbulent viscosity. Thus, we deduce that the inferred boost in $\Delta t$ required to explain TIC 241249530 b's properties may actually reflect the need for efficient dissipation by IWs. Given this, the constant $\Delta t$ framework we have utilized in this work is not strictly correct, but we can view it simply as a means of parameterizing the dissipation, which may actually come from IWs. Full coupled modeling of IW dissipation in highly eccentric giant planet formation is beyond the scope of this work but recommended for future work.

\section{HEM with Chaotic Dynamical Tides}
\label{sec: DT}

\subsection{Mathematical description}
\label{sec: DT math}
We now explore the chaotic dynamical tides model as a possible alternative to equilibrium tides in the secular evolution of the proto-WJ. Gaseous planets on highly eccentric orbits can have vibrational modes excited to large amplitudes that behave chaotically. We follow the treatment of \cite{vick2019chaotic}, which \cite{2025ApJ...989...35L} also recently added into the \texttt{REBOUNDx} library \citep{2020MNRAS.491.2885T}. The critical aspects of the mathematical model are reviewed below. 

We model the fluid composition of the planet as a $\gamma = 2$ polytrope and consider the evolution of the fundamental mode (f-mode), which is the most likely to be excited as a result of tidal forcing \citep{vick2019chaotic}. We let $\sigma$ denote the frequency of the f-mode in the inertial frame, given by \cite{vick2019chaotic} as
\begin{equation}
    \sigma = 1.22 \times \left(\frac{GM_p}{R_p^3}\right)^{1/2} + \Omega_s,
\end{equation}
where $\Omega_s$ is the rotation rate of the planet. 

During each pericenter passage, the amplitude of the f-mode is updated according to the hydrodynamic prescription of \cite{vick2019chaotic}. Specifically, the f-mode amplitude after the $k$\textsuperscript{th} pericenter passage is denoted $c_k$. The normalized mode amplitude $\tilde{c}_k$ is defined such that the mode energy at the $k$\textsuperscript{th} pericenter passage is $E_k = E_{B,0} |\tc_k|^2$, where $E_{B,0}$ is the initial orbital energy of the planet.

The parameter $\Delta E$ corresponds to the change in mode energy at periapse assuming $\tc = 0$ and is given by
\begin{equation}\label{eq:dE}
    \Delta E = \frac{GM_p^2}{r_p^6} R_p^5 T(\eta, \sigma, e)
\end{equation}
with $G$ the gravitational constant and $r_p$ the pericenter distance. The unitless function $T$ is given by \cite{vick2019chaotic} and depends the pericenter distance (in units of tidal radius) $\eta = r_p / r_{\text{tide}}$, the orbital eccentricity $e$, and the mode frequency $\sigma$. The tidal radius $r_\text{tide}$ is defined as 
\begin{equation}
    r_\text{tide} = \left(\frac{M_\star}{M_p}\right)^{1/3} R_p.
\end{equation}

At each pericenter passage, the change in the mode amplitude is given by $\Delta \tc = \sqrt{\Delta E / E_{B,0}}$, corresponding to an energy transfer of
\begin{equation}\label{eq:dEk}
    \Delta E_k = E_{B,0}(|\tc_{k-1} + \Delta\tc|^2 - |\tc_{k-1}|^2).
\end{equation}
The mode amplitude becomes
\begin{equation}\label{eq:tck}
    \tc_k = (\tc_{k-1} + \Delta \tc) e^{-i \sigma P_k},
\end{equation}
for $P_k$, the orbital period after the $k$\textsuperscript{th} pericenter passage.

A key parameter to determine whether this map will lead to chaotic behavior is the change in the phase of the f-mode after the $k$\textsuperscript{th} pericenter passage, which is denoted by $|\Delta \hat{P}|$. When $|\Delta \hat{P}| > \Delta \hat{P}_{\text{crit}} \sim 1$, the phase of the mode at subsequent pericenter passages is nearly random, leading to chaotic evolution of the planet's f-mode amplitude, and consequently, its orbit.

The parameter $|\Delta \hat{P}_k|$ is given by \cite{vick2019chaotic} as
\begin{align} \label{eq:dPhat}
    |\Delta \hat{P}| &= \sigma \Delta P\\
    &\simeq \frac{6\pi \sigma / \Omega_s}{(1-e)^{5/2}}  \left(\frac{M_p}{M_\star}\right)^{2/3} \eta^{-5} T(\eta, \sigma, e).
\end{align}
The condition $|\Delta \hat{P}_k| \gtrsim 1$ can be recast in terms of the planet's pericenter distance. Specifically, chaotic evolution occurs whenever $r_p \lesssim r_{p, \text{crit}}$ for
\begin{align}\label{eq:rpcrit}
    r_{p, \text{crit}} &\simeq (0.206 \text{ au}) \,\, \overline{\sigma}^{-0.59} \, Q^{0.11} \left(\frac{1-e}{0.02}\right)^{-0.135} \notag \\
    &\times  \left(\frac{R_p}{R_J}\right)\left(\frac{10^3 M_p}{M_\star}\right)^{-0.297} \left(\frac{\sigma}{\epsilon \Delta \hat{P}_{\text{crit}}}\right)^{0.054},
\end{align}
where $\overline{\sigma} = \sigma \left(GM_p/{R_p^3}\right)^{-1/2}$, $\epsilon$ has units of frequency and is strongly related to the mode frequency, and ${Q=0.56}$ for the f-mode. We note that non-linear effects may enhance chaotic tidal dissipation by decreasing $|\Delta \hat{P}_k|$ by a factor of $\sim 5$ \citep{yu2021tides}. However, since $r_{p, \text{crit}} \sim |\Delta \hat{P}_k|^{-0.054}$, this only amounts to a change in critical pericenter distance by $\sim 9\%$, which we neglect in the subsequent analysis.

In the context of vZLK-driven chaotic tidal migration, planets typically form with $r_p \gg r_{p, \text{crit}}$ and are driven to a high eccentricity and low pericenter distance through secular interactions with the perturbing body. If the planet's pericenter distance is driven to a value $\sim r_{p,\text{crit}}$, the subsequent evolution will be dominated by chaotic tides, which conserve angular momentum. At high eccentricities, the reduced angular momentum is $j^2 = a(1-e^2) \approx 2r_p$, meaning that the pericenter distance will remain $\sim r_{p,\text{crit}}$ for the duration of the chaotic evolution. This allows us to predict the post-chaotic migration pericenter distance of the planet analytically as $r_{p, \text{final}} \approx r_{p,\text{crit}}$. In reality, planets often undergo several phases of chaotic dissipation before vZLK oscillations are quenched. In this case, we have $r_{p, \text{final}} \approx r_{p,\text{crit}}$ during the final episode of chaotic migration. An equivalent way to denote this is $j^2_{\text{final}} \approx j_{\text{crit}}^2$. In Appendix \ref{sec: numerical validation}, we numerically validate this criterion using simulations in \texttt{REBOUND} \citep{rein2012rebound, rein2015ias15}, confirming that $j^2_{\text{final}} \approx j_{\text{crit}}^2$ accurately predicts the system configuration immediately following chaotic migration.

After the vZLK oscillations are quenched, the resulting dynamics are governed entirely by ETs and general relativity, which conserve angular momentum. Thus, we can constrain the post-chaotic evolution with the criterion
\begin{equation}
    j^2 \approx j_{\text{crit}}^2.
\end{equation}
Any planet whose orbital elements do not satisfy this criterion are not compatible with chaotic migration, unless the planet underwent substantial structural changes that implied a different $R_p$ and f-mode frequency, $\sigma$ (and thus a different $r_{p,\text{crit}}$), during the time of chaotic migration. We explore this possibility in Section \ref{sec:inflation}. 

\subsection{Application to TIC~241249530~b's formation}
\label{sec: secular integrations}

We now study the formation of TIC~241249530~b in the CDT framework. First we can consider the analytic criterion outlined in the previous section. We compare $j_{\text{obs}}^2 = 0.074\, \text{au}$ to the analytic $j^2_{p,\text{crit}} = 2r_{p,\text{crit}}$, calculated from equation \ref{eq:rpcrit} to be $j^2_{\text{crit}} \approx 0.026 \, \text{au}$, where we take $e=0.99$ and neglect $(\sigma/\epsilon \Delta \hat{P}_{\text{crit}})^{0.054}$ which is $\sim 1$ to a good approximation. The fact that $j_{\text{obs}}^2 = 0.074\, \text{au}$ is significantly greater than $j^2_{\text{crit}}$ already indicates a problem; if the planet migrated via chaotic tides, it should have been left with $j^2 \approx j^2_{\text{crit}}$. The mismatch suggests the planet was not emplaced by chaotic tides.  

We can further strengthen this argument with simulations. We modify \texttt{SecularMultiple} to include the effects of CDTs via the prescription of \cite{vick2019chaotic} and \cite{liveoak2025self}. Specifically, when $|\Delta \hat{P}|>1$, we set the integration timestep to the planet's orbital period. At each timestep with $|\Delta \hat{P}|>1$, we update $a$ and $e$ due the energy transfer between the f-mode and planet's orbit assuming angular momentum is conserved, via
\begin{align}
    e'&= \left(1 - \frac{E_B-\Delta E_k}{E_B}(1-e^2)\right)^{1/2}\\
    a' &= \frac{E_B}{E_B - \Delta E_k}a,
\end{align}
where $E_B$ is the orbital energy at the end of the previous step and $\Delta E_k $ is computed via equation \eqref{eq:dEk}.

In our simulations, we assume the proto-WJ has a pseudo-synchronous rotation rate given by 
\begin{equation}
    \Omega_s(e) = \frac{f_2(e)}{(1-e^2)^{3/2}} f_5(e)\, n
\end{equation}
where $n$ is the mean motion of the planet and $f_2(e)$ and $f_5(e)$ are given by
\begin{align}
    f_2(e) &= 1 + \frac{15}{2} e^2 + \frac{45}{8} e^4 + \frac{5}{16}e^6\\
    f_5(e) &= 1 + 3 e^2 + \frac{3}{8}e^4.
\end{align}
During the evolution, if the mode energy exceeds $E_{\text{max}} = 0.1E_{\text{bind}}$ where $E_{\text{bind}}^2 = GM_p/R_p$, we assume it is rapidly dissipated within a single orbit to ${E_{\text{resid}}=0.01 E_{\text{max}}}$. As discussed in \cite{vick2019chaotic}, the planet's orbit post-migration does not depend strongly on the particular choice of $E_{\text{max}}$ and $E_{\text{resid}}$. 

We initialize each simulation with TIC~241249530~b's parameters assigned as follows: we set $e_1=0.1$, and uniformly sample $a_1 \in [12\, \text{au}, 17\, \text{au}]$. We choose the argument of periapsis $\omega_1 =0 \degree$ and the longitude of the ascending node $\Omega = 0 \degree$. We assume that TIC~241249530~b has fixed radius $R=1.2\, R_J$. We include the effects of ETs by specifying tidal Love number $k_2=0.25$, tidal gyration radius $r_g=0.25$, and constant time lag $\Delta t=1 \, \text{second}$, though we note that these are only relevant for the post-chaotic migration phase. We initialize the binary companion on a circular orbit with $a_2 = 1664\,\text{au}$ and mutual inclination sampled uniformly from $[88\degree, 89 \degree]$. We choose the argument of periapsis $\omega_2 = 0 \degree$ and the longitude of the ascending node $\Omega_2 = 0 \degree$. 

\begin{figure}
    \centering
    \includegraphics[width=\linewidth]{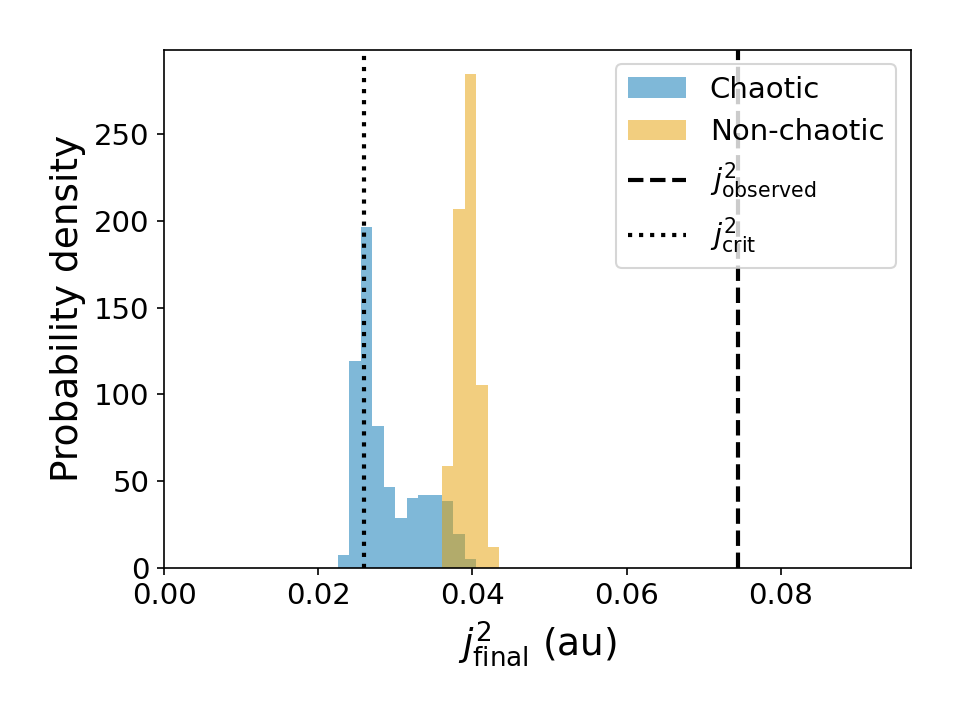}
    \caption{Distribution of final reduced angular momentum $j^2_{\text{final}}$ for systems simulated with \texttt{SecularMultiple} with DTs. The blue distribution indicates systems for which underwent chaotic migration ($\max|\Delta \hat{P}| > 1$), and the orange distribution indicates systems that migrated via ETs ($\max|\Delta \hat{P}| < 1$). The vertical dotted line indicates $j^2_{\text{crit}}$, where the chaotic values peak. The vertical dashed line indicates the observed reduced angular momentum of TIC~241249530~b, $j_{\text{obs}}^2 = 0.074\, \text{au}$, much larger than $j^2_{\text{crit}}$.} 
    \label{fig:dt-secular}
\end{figure}

We carry out 1000 integrations of these initial conditions, each for $1$ Gyr. Planets that completed migration such that the precession due to vZLK was quenched have their $j^2_{\text{final}}$ values plotted in \Cref{fig:dt-secular}. As expected, among systems that underwent chaotic tidal migration, the distribution is peaked near $j^2_{\text{final}} \approx j^2_\text{crit} \approx 0.026\, \text{au}$. This supports the conclusion drawn from the analytic criterion alone. Furthermore, there are some systems in this collection that did not reach small enough pericenter distances for chaotic migration to ensue and rather migrated via non-chaotic DTs and ETs. The distribution for these systems is peaked round $j^2 \approx 0.04\, \text{au} < j_{\text{obs}}^2$, due to the substantially weaker tidal dissipation compared to Section \ref{sec:ET} $(\Delta t = 1 \text{ second} $ vs. $\Delta t\sim 30 \text { seconds}$). Overall, all systems have $j^2_{\text{final}} < j^2_{\text{obs}}$ by a factor of a few.  Therefore, we conclude that under these circumstances TIC~241249530~b is not consistent with a history of chaotic tidal migration.

\section{Discussion}
\label{sec: discussion}

\subsection{Radius inflation cannot make CDTs compatible with TIC~241249530~b}
\label{sec:inflation}

While the mismatch between $j^2_{\text{obs}}$ and $j_{\text{crit}}^2$ is problematic, this is not necessarily the end for the chaotic tides hypothesis. It is evident from equation \ref{eq:rpcrit} that $r_{p,\text{crit}}$ depends on the planet's structure via the factors of $R_p$ and the f-mode frequency $\bar\sigma$. If tidal heating is significant from the CDTs, this could theoretically inflate the planet, leading to a different $r_{p,\text{crit}}$. Here we investigate whether tidally induced radius inflation could alter the planet's structure enough to rectify the mismatch with chaotic tides.   

\begin{figure}
    \centering
    \includegraphics[width=\linewidth]{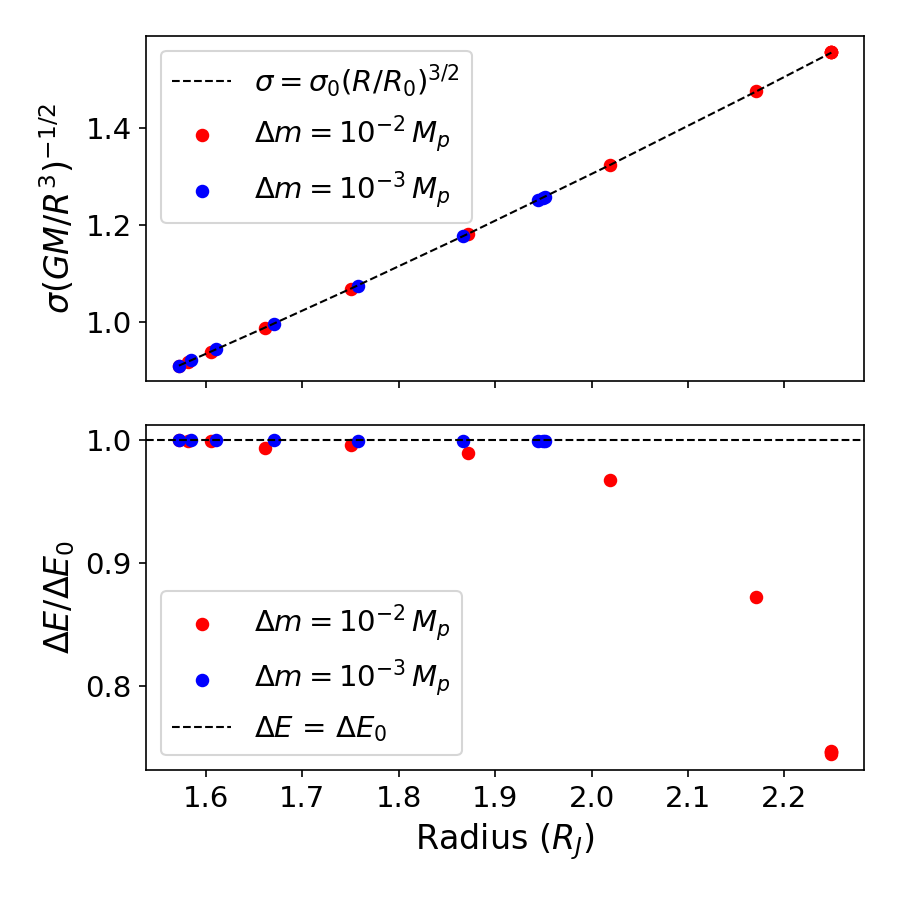}
    \caption{The f-mode frequency and tidal energy transfer at pericenter for a giant planet that is inflating due to uniform heating in an outer envelope of depth $\Delta m$. The giant planet is modeled using \texttt{MESA} and the mode properties are determined with \texttt{GYRE}. The heating rate is $3.5\times 10^{30}~\rm {erg/s}$. The red and blue points show results from heating at two different depths, $\Delta m = 0.01 M_p$ and $\Delta m = 0.001 M_p$. The top panel shows $\sigma$ in units where $G=M_p=R_p=1$ and the bottom panel shows $\Delta E$ from equation~(\ref{eq:dE}) calculated for $a= 2~$AU and $e=0.98$, scaled to the value for an unheated planet, $\Delta E_0$. Dashed lines correspond to constant $\sigma$ and $\Delta E$ as the planet inflates.}
    \label{fig:inflation}
\end{figure}

    For a given orbit, the threshold for chaotic tidal migration depends on the structure of the planet. According to  equation~(\ref{eq:rpcrit}), it seems that a tidally inflated planet, with increased $R_p$, would have a larger $r_{p, \rm crit}$ and could migrate more easily. However, as the planet's structure evolves, the mode parameters $\bar{\sigma}$ and $Q$, both normalized such that $R_p=1$, also change. We must understand how the mode properties of a planet are affected by tidal heating to understand whether an inflated planet truly is more vulnerable to chaotic migration.

    Chaotic tidal heating likely deposits energy in the outermost layers of the planet. This is the region where non-linear effects like wave-breaking can dissipate energy because the f-mode oscillations are largest near the surface of the planet \citep{2018AJ....155..118W}. We model chaotic tidal heating as occurring uniformly by mass in the outer envelope of the planet with a depth of mass $\Delta m$, and use the \texttt{MESA} stellar evolution code (version 10108) to compute the structural evolution of a $1 M_J$ giant planet as it is heated \citep{Paxton2011, Paxton2013, Paxton2015, Paxton2018, Paxton2019, Jermyn2023}. We adopted a tidal heating rate of $3.5 \times 10^{30}~\rm{erg/s}$, which corresponds to a change in the semimajor axis from $\sim1.5$~AU to 0.5~AU over the course of $10^5$ years for a $1M_J$ planet orbiting a $1M_\odot$ star -- typical values for a giant planet undergoing chaotic tidal migration. The \texttt{GYRE} asteroseismology code was used to evaluate $\bar{\sigma}$ and $Q$ for the inflating giant planet \citep{Townsend2013}. 

    The results of these calculations are shown in Figure~\ref{fig:inflation}. The top panel is the f-mode frequency ${\bar{\sigma} = \sigma(GM_p/R_p^3)^{-1/2}}$ plotted against the radius of the planet as it expands to a new equilibrium value due to tidal heating. For both heating depths, $\Delta m = 0.01 M_p$ and  $0.001 M_p$, we find that $\bar{\sigma} \propto R_p^{3/2}$. This corresponds to a constant value of $\sigma$ while the planet inflates. The bottom panel shows $\Delta E$ from equation~(\ref{eq:dE}), calculated for a heated planet at an orbit with $a= 2~$AU and $e=0.98$, scaled to $\Delta E_0$ calculated for an unheated planet. For a heating depth of $\Delta m = 0.01 M_p,$ energy transfer actually decreases for a planet that is inflated due to chaotic tidal heating. For shallower heating with $\Delta m = 0.001 M_p,$ energy transfer is nearly constant. 

    For heating that occurs in a shallow envelope, the tidal energy transfer at pericenter is relatively unchanged, even while the radius of the planet increases significantly. Although the planet is inflated, its structure is largely the same, except for very close to the planet surface. As a result, the f-mode properties for a heated planet are very similar to those of an unheated planet, and tidal energy transfer does not increase with the planet radius. Therefore, we do not expect this type of tidal heating to increase $r_{p,{\rm crit}}$ and make giant planets more vulnerable to chaotic tidal migration.

    This result is in agreement with recent work by \cite{2026arXiv260620789Z}, who studied the coupled orbital and structural evolution of giant planets undergoing high-eccentricity migration via dynamical tides. While they showed that very tight periastron passages can drive hydrodynamic winds that unbind large fractions of the planetary atmosphere, larger periastron distances are associated with mode energies that either stagnate or grow chaotically with shallow shocks that are damped by radiative diffusion. \cite{2026arXiv260620789Z} showed that TIC 241249530 b currently lies in the stagnant regime, with a pericenter distance too large to chaotically grow the f-mode energies, consistent with our findings.

\subsection{Mass loss cannot make CDTs compatible with TIC 241249530 b}
Distinct from near-total atmosphere loss, giant planets that reach very eccentric orbits can be subject to mass loss via partial tidal disruption \citep{guillochon2011consequences, yu2026dynamical, weldon2026saving}. Recently, \cite{weldon2026saving} conducted a comprehensive study of the role of mass loss in giant planets experiencing vZLK-driven high-eccentricity migration. They found that giant planets experience mass loss at around 5\% on average, although sometimes much larger, and the mass transfer process can return angular momentum to the orbit. If TIC 241249530 b underwent mass loss during a history of chaotic tidal migration, the final specific angular momentum $j_{\text{final}}^2$ may be higher than $j_{\text{crit}}^2$, thus potentially relieving the tension between $j_{\text{obs}}^2$ and $j_{\text{crit}}^2$. We briefly evaluate this possibility here.

For TIC 241249530 b to arrive at its current angular momentum in the case of partial disruption during chaotic tidal migration, it would require
\begin{equation}
    \frac{m_{i}}{m_{\text{obs}}}=\frac{j_{\text{obs}}}{j_{\text{crit}}} \sim 1.5-2,
\end{equation}
where $m_i$ is the initial mass and this assumes the maximal fraction of angular momentum return to the orbit \citep{weldon2026saving}. According to \cite{weldon2026saving}’s simulations, planets are not expected to lose this much mass if they are to end as massive as $\sim5 \ M_{\mathrm{Jup}}$. Thus, we conclude that partial tidal disruption during chaotic tidal evolution is unlikely to explain TIC 241249530 b's present-day orbit. However, we recommend that a future meta-analysis of all highly eccentric hot Jupiter progenitors explore this possibility.

\subsection{Consequences for observed HJs and WJs}
\begin{figure*}
    \centering
    \includegraphics[width=\linewidth]{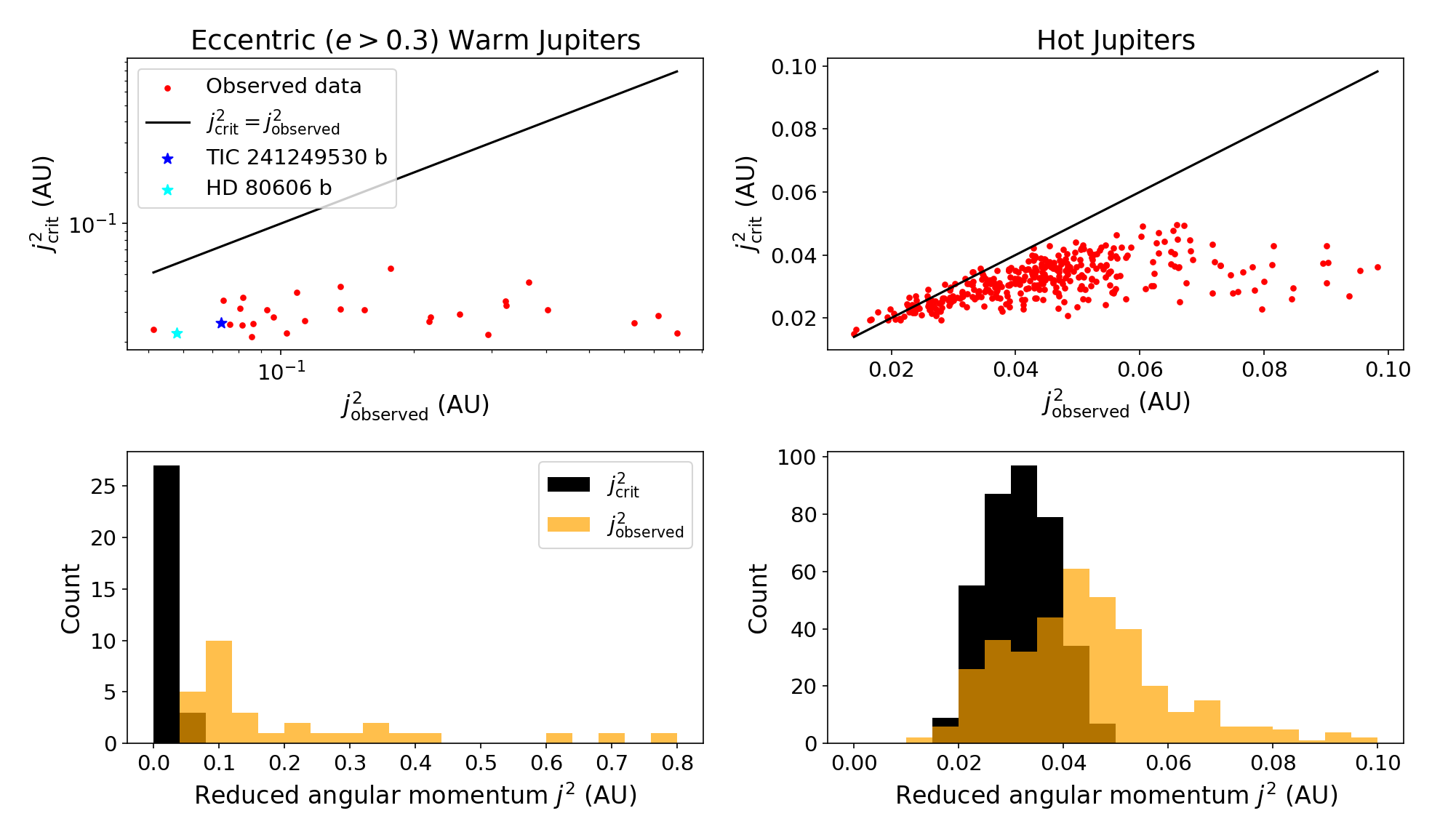}
    \caption{Top panels: $j^2_{\text{crit}}$ vs. $j^2_{\text{obs}}$ for observed eccentric WJs $(e > 0.3)$ and HJs. A black line indicates $j^2_{\text{crit}} = j^2_{\text{obs}}$, and TIC~241249530~b and HD 80606 b are indicated with dark blue and cyan stars, respectively. Bottom panels: distribution of $j^2_{\text{crit}}$ (black) and $j^2_{\text{obs}}$ (orange) for observed eccentric WJs and HJs.}
    \label{fig:observedHJWJ}
\end{figure*}
Even if chaotic tidal migration appears unlikely for TIC~241249530~b, the analytic criterion $j^2_{\text{final}} \approx j^2_\text{crit}$ gives us an opportunity to investigate the broader population of observed HJs and eccentric WJs to determine if we can see evidence of a history of chaotic tidal migration. We use the NASA Exoplanet Archive \citep{2013PASP..125..989A, christiansen2025nasa, https://doi.org/10.26133/nea1} and limit our analysis to planets with well-constrained masses, semi-major axes, and eccentricities. 

We classify a planet as a HJ if it has mass between $0.5\, M_J$ and $7 \, M_J$, radius above $0.8\, R_J$, orbital eccentricity $e<0.1$ and semi-major axis $a<0.1\, \text{au}$. We clasify a planet as an eccentric WJ if it has mass between $0.5\, M_J$ and $7 \, M_J$, radius above $0.8\, R_J$, orbital eccentricity $e>0.3$ and semi-major axis between $0.1\, \text{au}$ and $1\,\text{au}$. For each planet, we compute $j^2_{\text{crit}}\approx 2 r_{p, \text{crit}}$ via equation \eqref{eq:rpcrit}, assuming a maximum eccentricity of $e=0.99$ and pseudo-synchronous rotation rate.

The results of this analysis are shown in \Cref{fig:observedHJWJ}. We note that all eccentric WJs have $j_{\text{obs}}^2 > j_{\text{crit}}^2$ by a factor of $\sim $ a few, as in the case of TIC~241249530~b, and thus are likely not compatible with chaotic tidal migration. Interestingly, the closest HJs $(a\lesssim 0.03\, \text{au})$ satisfy $j_{\text{obs}}^2 \approx j_{\text{crit}}^2$. On one hand, this would be the expected signature of chaotic migration, but on the other, the agreement could merely be a coincidence. Indeed, prior studies of HEM with ETs also correctly reproduced the observed HJ pile-up at $0.03-0.06$ au \citep{naoz2012formation, 2015ApJ...805...75P, 2016MNRAS.456.3671A}. Further analysis, including detailed modeling of individual systems (especially those with distant companions), would be required to evaluate whether or not our observed trend is evidence of chaotic migration.

\section{Conclusion}
Highly eccentric hot Jupiter progenitors are the most direct evidence of high-eccentricity migration, and they offer valuable testbeds of the detailed physics of the mechanism. Here we explored the recently discovered TIC~241249530~b and constrained its migration history. We separately considered equilibrium tides and chaotic dynamical tides and explored which scenario creates model planets most resembling TIC~241249530~b. Our conclusions are as follows.
\begin{enumerate}
\item TIC~241249530~b's observed properties are well explained by von Zeipel-Lidov-Kozai cycles triggered by the observed distant binary star and dissipated via equilibrium tides, but only with a time lag $\Delta t \gtrsim 30 \text{ seconds}$, about $300\times $ times more dissipative than Jupiter. This may reflect efficient dissipation by inertial waves, and we recommend future study of the role of inertial waves in eccentric giant planet formation.
\item We identify a simple analytic criterion pertinent to chaotic migration: the post-migration value of $j^2=a(1-e^2)$ is  $j^2=j^2_{\text{crit}} \approx 2 r_{p,\text{crit}}$, where $r_{p,\text{crit}}$ (equation \ref{eq:rpcrit}) is the critical pericenter distance for chaotic migration.
\item Using both the analytic criterion and simulations of chaotic dynamical tides, we find $j^2_{\text{obs}} > j^2_{\text{crit}}$ by a factor of a few, suggesting that TIC~241249530~b is not consistent with chaotic migration. 
\item This inconsistency cannot be rectified by planetary structure changes due to heating as a result of chaotic tidal dissipation, nor mass loss due to partial tidal disruption.
\item We analyze the comparison between $j^2_{\text{obs}}$ and $j^2_{\text{crit}}$ for the observed population of hot Jupiters and eccentric warm Jupiters. For WJs, $j^2_{\text{obs}}$ > $j^2_{\text{crit}}$, but many close-in hot Jupiters have $j^2_{\text{obs}} \approx j^2_{\text{crit}}$. This is a tantalizing possible signature of chaotic migration, but it may also be coincidental; further modeling of individual systems is recommended. 
\end{enumerate}

\section{Acknowledgments} We gratefully acknowledge the anonymous referee, whose comments substantially improved our paper. This material is based upon work supported by the National Science Foundation under grant No. 2306391. We also acknowledge support from the MIT Undergraduate Research Opportunities Program. We gratefully acknowledge access to computational resources through the MIT Engaging cluster at the Massachusetts Green High Performance Computing Center (MGHPCC) facility and the MIT SuperCloud and Lincoln Laboratory Supercomputing Center \citep{reuther2018interactive}. This research has made use of the NASA Exoplanet Archive, which is operated by the California Institute of Technology, under contract with the National Aeronautics and Space Administration under the Exoplanet Exploration Program.

\appendix
\section{Numerical validation of chaotic migration criterion}
\label{sec: numerical validation}

To confirm the analytic criterion presented in Section \ref{sec: DT math}, we carry out $N$-body integrations using the \texttt{REBOUND} package \citep{rein2012rebound}. We use the \texttt{tides\_dynamical} \citep{liveoak2025self} effect implemented in \texttt{REBOUNDx} \citep{tamayo2020reboundx} to model chaotic tidal migration. We use the adaptive timestep integrator \texttt{IAS15} \citep{rein2015ias15} to ensure precision despite high eccentricities.

We construct two suites of 1000 integrations each, one with proto-WJ mass $1\, M_J$ and the other with proto-WJ mass $5\, M_J$. In both suites, we uniformly sample the planet radius from $1$ to $3$ $R_J$. Note that the parameters chosen in these simulations correspond to a generalized system -- not TIC 241249530 -- since our goal is merely to validate the analytic criterion. In each simulation, the planet is initialized on a circular orbit around a star with mass $M_\star = 1\, M_\odot$. A binary companion of mass $M_b = 1\, M_\odot$ is initialized on a circular orbit with a mutual inclination of $i_0 = 87\degree$ from the proto-WJ. The semi-major axes of the proto-WJ and binary companion are selected uniformly from $[3\, \text{au},5\, \text{au}]$ and $[300\,\text{au}, 800\, \text{au}]$ respectively. We integrate each system for $500 \, \text{Myr}$, or until the apisidal precession due to SRFs exceeds $10 \times$ that due to the vZLK mechanism \citep{liu2015suppression}.

\begin{figure}
    \centering
    \includegraphics[width=\linewidth]{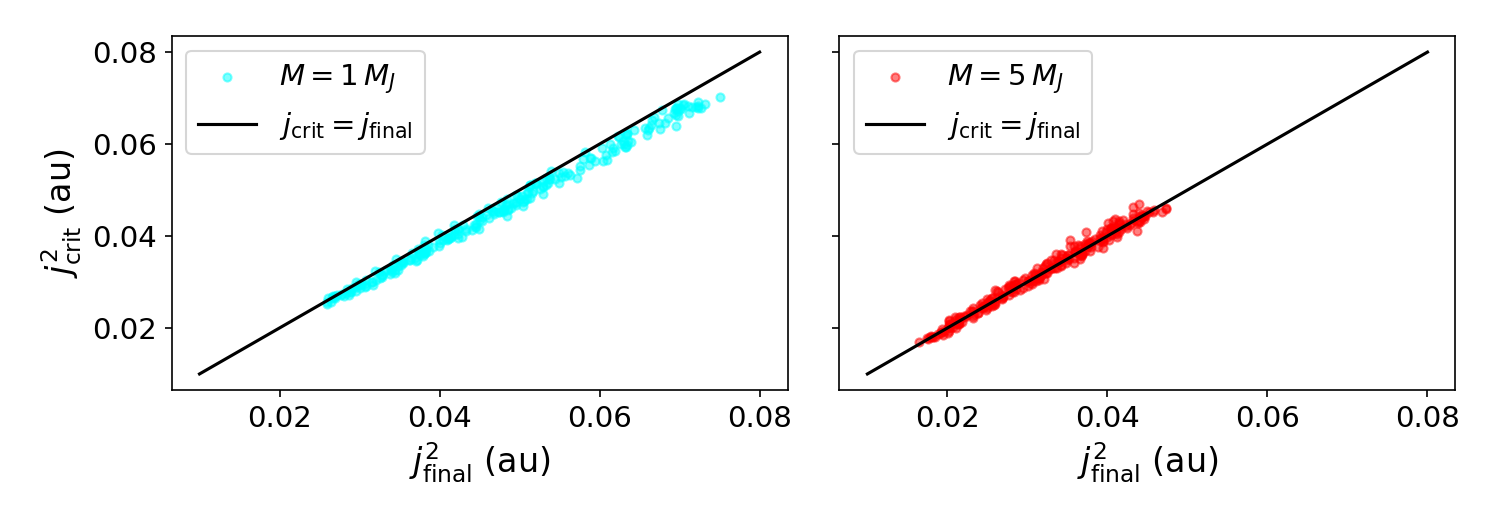}
    \caption{$j^2_{\text{crit}}$ vs. $j_{\text{final}}^2$ for the $1\, M_J$ (left) and $5\, M_J$ (right) suites of $N$-body integrations. The black lines indicate $j^2_{\text{final}} = j^2_{\text{crit}}$.}
    \label{fig:reboundx-sims}
\end{figure}

In the $1\, M_J$ and $5\, M_J$ samples, 254 and 267 systems underwent chaotic tidal migration to the point that the vZLK oscillations were quenched, respectively. For each of these systems, we compute $j^2_{\text{crit}}=2 r_{p, \text{crit}}$ assuming a pseudo-synchronous rotation rate with $e=e_{\text{max}}$, the maximum eccentricity during the integration. \Cref{fig:reboundx-sims} shows $j^2_{\text{crit}}$ vs. $j_{\text{final}}^2$ for each system. We see that the approximation $j^2_{\text{final}} \approx j_{\text{crit}}^2$ holds to a high degree of accuracy for both integration suites.

\bibliographystyle{aasjournal}
\bibliography{main}

\end{document}